\def\ta2o5{$\mathrm{Ta_2O_5}$}
\def\si2n3{$\mathrm{Si_2N_3}$}

\newcommand{\degr}{\ensuremath{^\circ\:}}

\documentclass [twocolumn,superscriptaddress,preprintnumbers]{revtex4}
\usepackage {graphicx}
\usepackage {amsmath}
\usepackage {bm}
\usepackage {subfigure}

\begin{document}

\title{
Three-dimensional coherent X-ray diffraction imaging of a ceramic
nanofoam: determination of structural deformation mechanisms
}

\begin{abstract}
Ultra-low density polymers, metals, and ceramic nanofoams are
valued for their high strength-to-weight ratio, high surface area
and insulating properties ascribed to their structural geometry. We
obtain the labrynthine internal structure of a tantalum oxide nanofoam
by X-ray diffractive imaging. Finite element analysis from the
structure reveals mechanical properties consistent with bulk samples
and with a diffusion limited cluster aggregation model, while excess
mass on the nodes discounts the dangling fragments hypothesis of
percolation theory. 
\end{abstract}

\author{A. Barty}
\affiliation{Lawrence Livermore National Laboratory, 7000 East Ave., Livermore, CA 94550, USA}
\author{S.~Marchesini}
\affiliation{Lawrence Livermore National Laboratory, 7000 East Ave.,
Livermore, CA 94550, USA}
\affiliation{Lawrence Berkeley National Laboratory, 1 Cyclotron
rd. Berkeley, CA 94720, USA}
\affiliation{Center for Biophotonics Science and Technology, University of California, Davis, 2700 Stockton Blvd., Ste 1400, Sacramento, CA 95817, USA}
\email[Correspondence and requests for materials should be addressed
to S. Marchesini:]{smarchesini@lbl.gov}

\author{H. N. Chapman}
\affiliation{Lawrence Livermore National Laboratory, 7000 East Ave., Livermore, CA 94550, USA}
\affiliation{Center for Biophotonics Science and Technology, University of California, Davis, 2700 Stockton Blvd., Ste 1400, Sacramento, CA 95817, USA}

\author{C. Cui}
\author{M. R. Howells}
\author{D.A.  Shapiro}
\author{A. M. Minor}
\affiliation{Lawrence Berkeley National Laboratory, 1 Cyclotron rd. Berkeley, CA 94720, USA}
\author{J. C. H. Spence}
\affiliation{Department of Physics and Astronomy, Arizona State University, Tempe, AZ 85287-1504, USA}
\author{U. Weierstall}
\affiliation{Department of Physics and Astronomy, Arizona State University, Tempe, AZ 85287-1504, USA}
\author{J. Ilavsky}  
\affiliation{Advanced Photon Source, Argonne National Laboratory,
Argonne, IL, USA}
\author{A. Noy}
\author{S. P. Hau-Riege} 
\author{A. B. Artyukhin}
\author{T. Baumann}
\author{T. Willey}
\author{J. Stolken} 
\author{T. van Buuren} 
\author{J. H. Kinney}
\affiliation{Lawrence Livermore National Laboratory, 7000 East Ave., Livermore, CA 94550, USA}


\preprint{UCRL-JRNL-231416}

\date{\today}
\maketitle

The topology, fractal index, stability and structure of foams have
fascinated scientists and mathematicians for decades. Foams arise in
fields as diverse as cosmology (in Hawking's theory), geology,
surfactants, phospholipids, cells, bone structure, polymers and
structural materials wherever lightness and strength are needed.
Especially important are applications of periodic foam network theory
to predictions of the structure of mesoporous crystalline materials
suitable for use as catalysts \cite{santanu} for cleaner fuels, and
the study of the diffusion of water and oil in porous rocks. Here the
rate-limiting step for diffusion is limited by the smallest pore,
normally too small to be observed internally by any conventional
tomographic microscopy.  Aerogels are an important example of such a
class of material.  Described variously as "frozen smoke" and ``San
Francisco fog", these terms do not refer to a particular substance
itself but rather to a structural geometry a substance can assume.
Many aerogels, for example, demonstrate astonishing mechanical,
thermal, catalytic and optical properties, which are ascribed to their
low density and porous structure\cite{hrubesh}.  However to date there
have been few if any methods developed for ``seeing inside" these
foams in order to make an experimental determination of topology and
structure at the mesoscopic length scale.

Although electron microscopy has provided the crystallinity and
morphology of individual beam elements comprising the foam, no
existing technique has been able to capture the three-dimensional bulk
lattice arrangement over micron-scale sample dimensions.  X-rays
provide the penetration, lacking to electrons, which allow us to study
three-dimensional structure over thicknesses of micrometres.  Current
state-of-the art zone plate microscopes achieve a transverse spatial
resolution of $\approx$15 nm half pitch (30 nm period) \cite{chao}, but have
difficulty sustaining this resolution through the bulk structure of
micron-sized three dimensional objects due to depth of focus
limitations.

We report here a three-dimensional structure
determination at the mesoporous length scale ($\approx$15 nm) of a
micron-sized fragment of aerogel obtained by inversion of coherent
x-ray diffraction patterns \cite{miao}. The complexity of the
structure observed is far greater than that of samples previously
studied by this technique, and was made possible by advances in
computational phase retrieval methods, the addition of holographic
reference points near the specimen, and by the inherent sparsity of
the foam.  More generally, we demonstrate the ability of diffraction
imaging to image an unknown, isolated object at high resolution in
three dimensions, opening the door to a wide range of applications in
material science, nanotechnology and cellular biology.

The aerogel sample imaged here is a low density (100 mg/cc, 1.2\% bulk
density) high-Z \ta2o5 metal oxide nanofoam, chosen because of its
potential use in double shell laser ignition targets for fusion (which
require a very low density material with high Z) and because of its
stability under an intense x-ray beam.  Because the strength of these
ultra-low density foams is orders of magnitude less than expected, it
is important to understand and identify the unique microstructures of
these foams, and to relate them to the bulk physical properties.
Using the structure observed as a high-fidelity template for
finite-element analysis we calculate the load displacement response of
the foam, and compare the resulting stiffness against various
structural models.

\begin{figure}
	\includegraphics[width=0.48\textwidth]{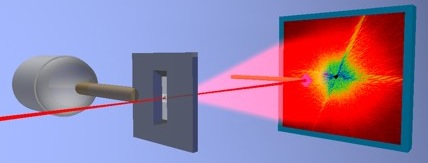}
\caption{
Diffraction imaging layout. Coherent X-rays ($\lambda=1.65$ nm)
 illuminate \cite{howells, beetz} the sample mounted on a 50 nm thick
\si2n3  membrane window. Diffraction patterns at various sample
orientations (+69\degr to -64\degr in  1$\pm 0.1\degr$ increments) 
are  measured using a CCD camera (20 $\mu$m pixel size,
1300$\times 1340$ pixels,  165 mm downstream). A beamstop blocks the
direct beam and multiple exposure times are summed to expand the 
CCD dynamic range. 
\label{fig:1}}
\end{figure}

X-ray diffraction imaging is elegant in its experimental simplicity: a
monochromatic and coherent X-ray beam illuminates the sample, and the
far-field diffraction pattern from the object is recorded on an area
detector (Figure \ref{fig:1}). Multiple orientations fill out a
three-dimensional diffraction volume, which is proportional to the
Fourier transform of the object index of refraction. The detection
system records only the diffracted intensities, but phase retrieval
techniques can be applied to recover a three-dimensional image of the
sample  \cite{fienup, luke, marchesini:rsi}.  The feasibility of
this technique for reconstructing an image of the sample from its
diffraction pattern has been well demonstrated in many X-ray
diffraction experiments \cite{miao, miao:prl, shapiro,
Chapman_Pyramid, pfeifer}.

A 3D implementation of the HIO algorithm \cite{fienup} was used for phase
reconstruction on the 3D volume with feedback parameter $\beta=0.9$ and support refinment
\cite{marchesini} for 1200 iterations, followed by the RAAR algorithm
\cite{luke} through to iteration 3000 also with feedback parameter of
$\beta=0.9$. Missing data from both the central beamstop and
inaccessible sample rotations was accounted for during phase retrieval
using the \emph{Shrinkwrap} algorithm, in which the object support
itself acts as a constraint in the regions of missing data
\cite{Chapman_Pyramid, marchesini}.  
By applying full three dimensional phase retrieval directly to the 3D
diffraction volume, as done here, we avoid the inconsistencies and
image alignment problems caused by dividing the phase retrieval and
tomographic data assembly into two separate steps performed
sequentially \cite{miao:prl}. Ewald sphere curvature included in the
3D data assembly avoids defocus artifacts typical of such two step
reconstruction process, as well as any lens-based micro-tomography
volume reconstruction.  The phase reconstruction techniques employed
here, including a detailed analysis of spatial resolution and methods
for handling the limited number of views, missing angles and central
beamstop are described in detail elsewhere \cite{Chapman_Pyramid} (see
also supplemental material\cite{supplemental}).

\begin{figure}[tbp]
\subfigure[]
    {
      \label{subfig:2a}
	\includegraphics[width=0.21\textwidth]{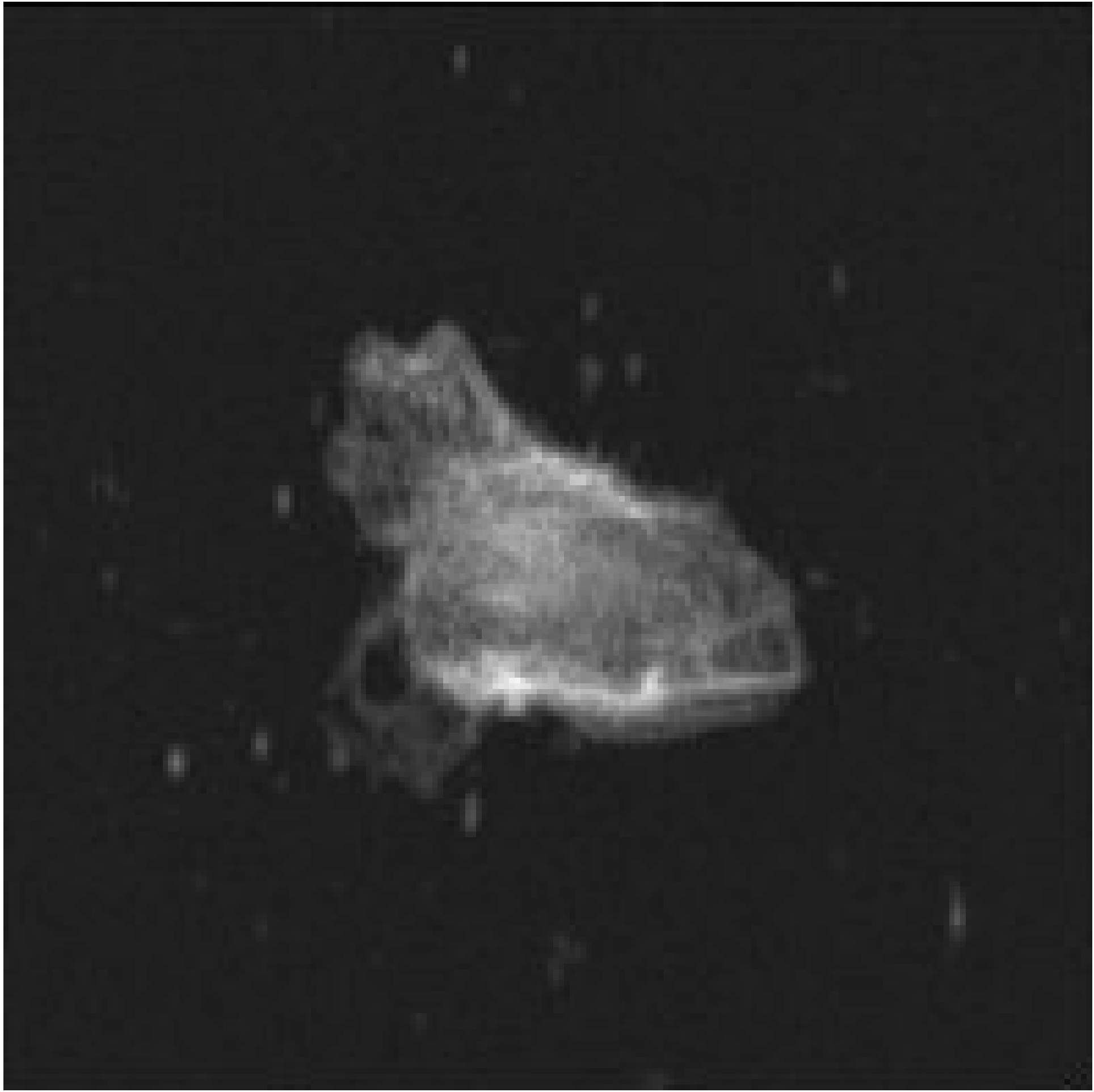}
    }
\subfigure[]
    {
      \label{subfig:2b}
	\includegraphics[width=0.21\textwidth]{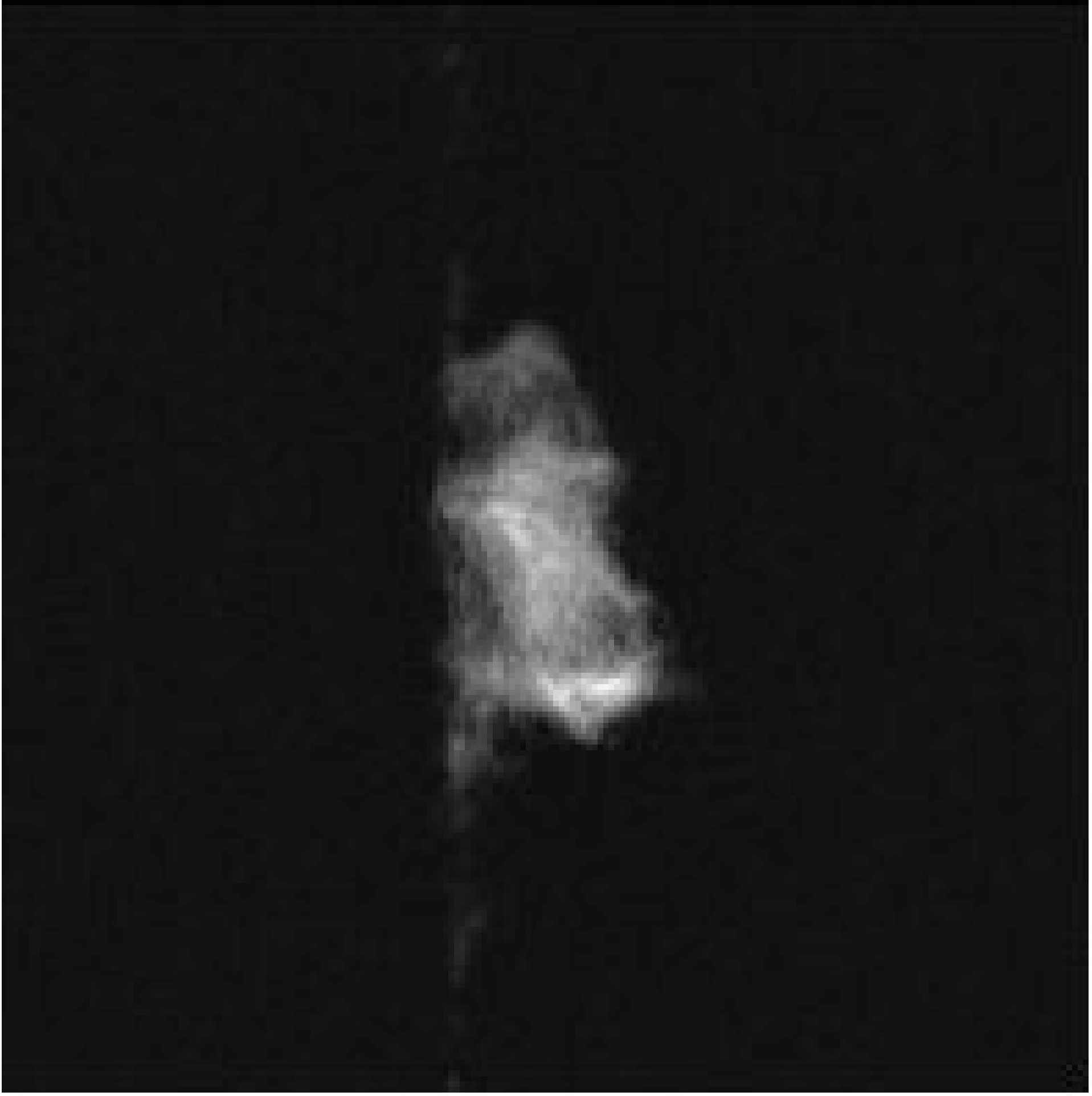}
    }\\
\subfigure[]
    {
      \label{subfig:2c}
	\includegraphics[width=0.21\textwidth]{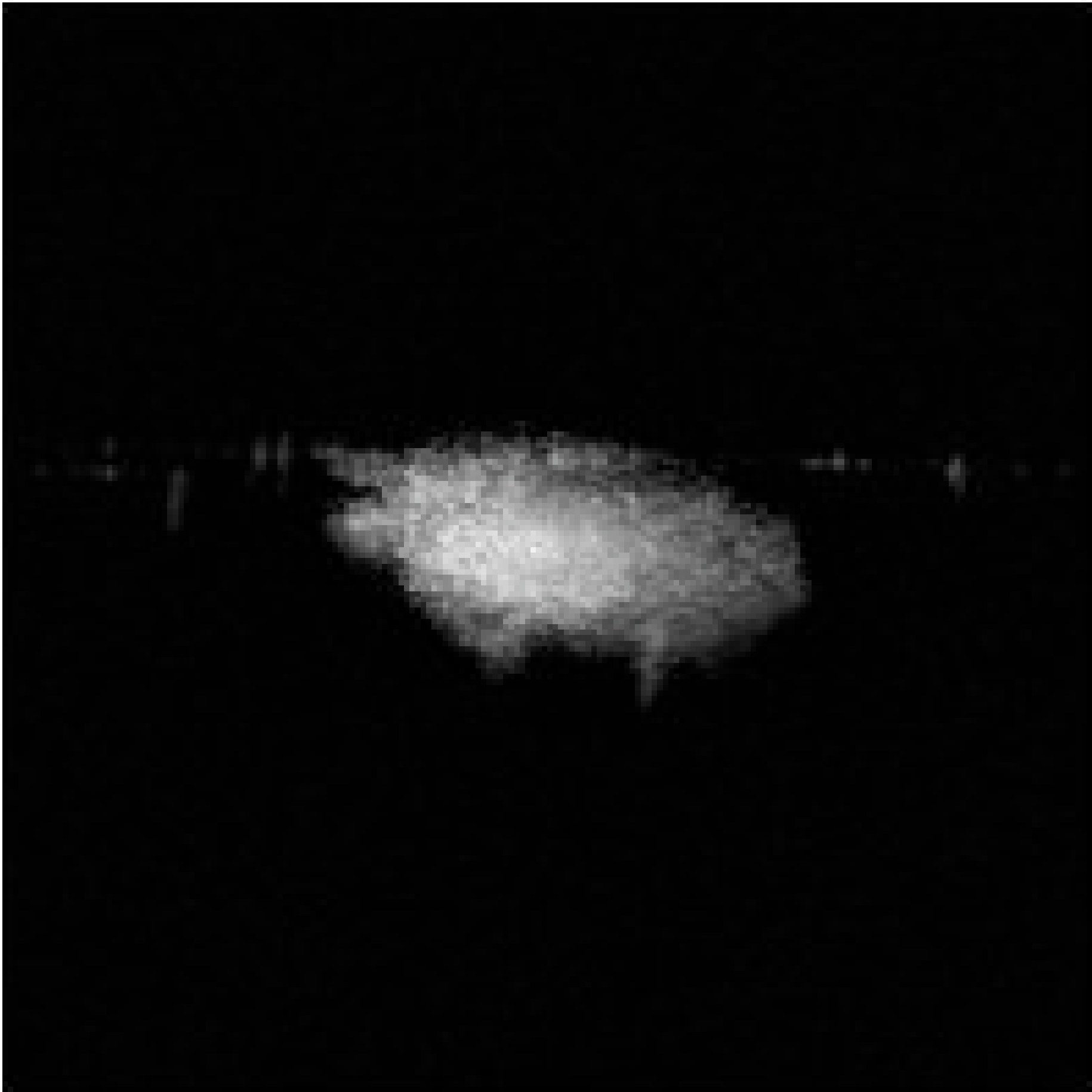}
    }
\subfigure[]
    {
      \label{subfig:2d}
	\includegraphics[width=0.21\textwidth]{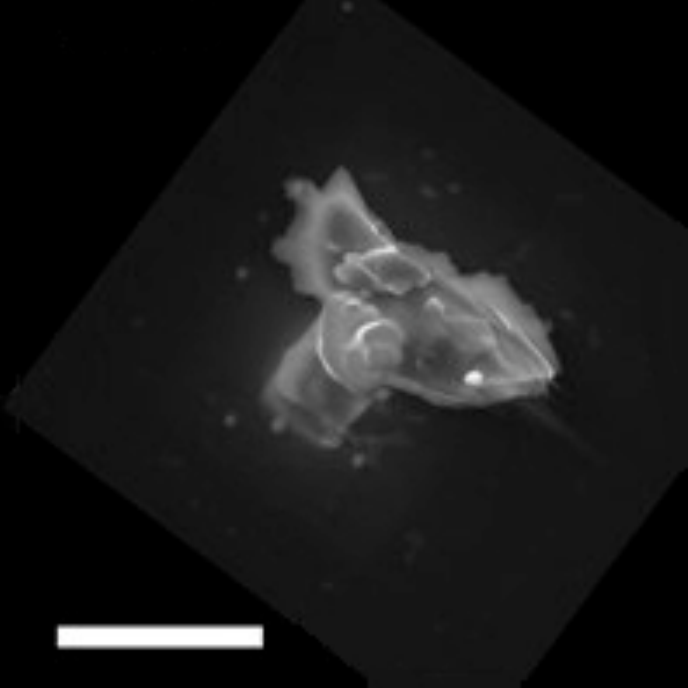}
    }
\caption{
(a-c) Orthogonal projections through the amplitude of the
reconstructed 3D object. The imaged volume is 5 $\mu$m cubed with a
reconstruction resolution of 15 nm. A top-down SEM image of the
prepared aerogel sample for comparison with the 3D reconstruction is
shown in panel (d), scale corresponds to 2 $\mu$m. 
Several reference platinum dots (d) are reconstructed on the membrane 
plane  (b, c) and the animation in
supplementary materials\cite{supplemental}.
\label{fig:2}}
\end{figure}

\begin{figure}
	\includegraphics[width=0.48\textwidth]{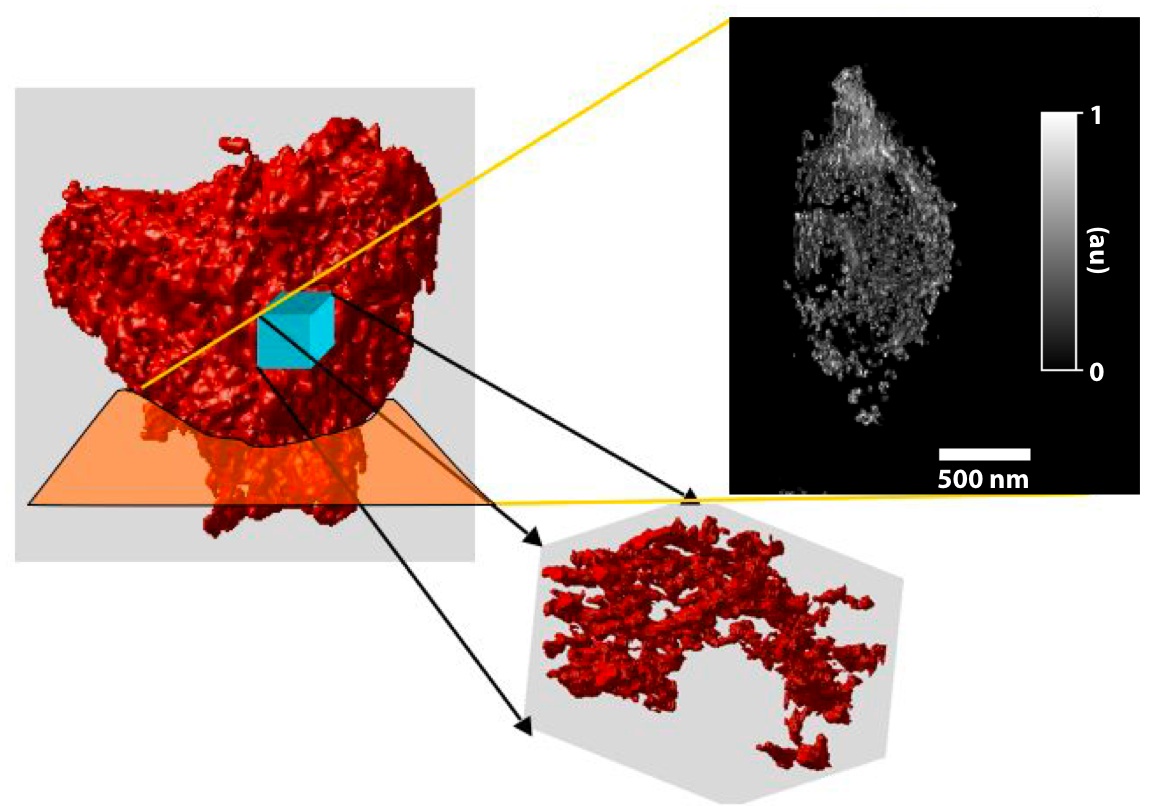}
\caption{
Section and isosurface rendering of a 500 nm cube 
from the interior of the 3D volume. The foam structure 
shows globular nodes that are interconnected by thin beam- 
like struts. Approximately 85 \% of the total mass is associated 
with the nodes, and there is no evidence of asignificant fraction 
of dangling fragments. 
\label{fig:3}
}
\end{figure}

The aerogel itself was prepared by a sol-gel process that involved the
controlled hydrolysis of tantalum ethoxide, followed by rapid
supercritical extraction of the reaction solvent \cite{brinker}. The
tantala aerogels obtained from this particular formulation are
isolated as translucent monoliths with bulk densities of approximately
100 mg/cc, as determined from bulk sample mass and dimensions.  We
mounted an irregular 1-2 $\mu$m sized piece on a 50 nm thick
rectangular silicon nitride membrane window of 2 mm x 50 $\mu$m size
supported in a 200 $\mu$m thick silicon wafer frame. Several 50 nm
platinum dots were placed in proximity to the sample using an electron
beam to locally decompose a metal-organic precursor gas (Figure
\ref{fig:2}). 
These dots diffract reference waves that provide holographic
information and act as heavy atoms in the phase retrieval process, and
additionally provide known structures on the Silicon membrane for
verification of reconstruction fidelity and spatial resolution (Figure
\ref{fig:3}).  High resolution structural information is required to
fully characterize mechanical properties of these aerogels, however
many other statistical properties such as density and correlation
distances can be obtained directly by standard analysis of the
radially averaged small angle x-ray scattering (SAXS) patterns.  We
note however that the radially averaged raw diffraction data used in
our reconstruction has a power law exponent of about -4, while
diffraction patterns generated by a reconstructed 3D region containing
only aerogel material reveals a power law exponent of -2
(Fig. \ref{fig:4}), typical of the ``string of pearls'' aerogel
morphology (see below).  This is because our measured data contains
additional scattering from the membrane, surrounding particles and Pt
contamination.  These scatterers are physically separated in the 3D
reconstruction, enabling scattering from just the aerogel portion of
the sample to be calculated.

X-ray scattering measurements extending the spatial scales probed by
diffractive imaging data were performed on a similar batch of
aerogel foam at beamline 33-ID at the Advanced Photon Source (APS)
\cite{USAXS_Ilavsky1, USAXS_Ilavsky2}. Three distinct power-law
regimes are observed at different spatial scales
(Fig. \ref{fig:4}). In the Porod region below q$\simeq 0.1$
\AA$^{-1}$ a power law with an exponent of -4 indicates that the foam
has a smooth surface.  The form factor region from q$\simeq 0.1$
\AA$^{-1}$ to q$\simeq 0.005$ \AA$^{-1}$ 1 provides information on the
shape of the individual scattering elements in the aerogel network. In
this region we have a power law exponent of -2, which describes a mass
fractal of dimension 2 which is consistent with a diffusion limited
cluster aggregation or the typical ``string of pearls'' aerogel
morphology.  Oscillations in the calculated (XDM) scattering are due
to the size of aerogel sub-region used to create the calculated plot.
The cross over points between the changing slopes occur at q$_1\simeq
0.1$ \AA$^{-1}$ and q$_2\simeq 0.09$ \AA$^{-1}$ with an associated
radii of gyration of $Rg_1\simeq\, 20 \AA$ and $Rg_2\simeq 140\, \AA$
respectively. Values of the power law exponents and radii
of gyration are obtained by standard Porod and Guinier analysis. For
aerogels a fractal analysis is commonly used to interpret the
scattering data. In the fractal model $Rg_1$ and $Rg_2$ are
related to the mean particle diameter and the correlation range, where the relation between $Rg$ and size
or correlation range depends strongly on the shapes and size
distribution of the scatters. At very low $q$ we find
another power law slope associated with another scattering feature,
however the USAXS data end before a clear radius of gyration is
observed therefore it is not possible to estimate the size of these
structures. We can, however, limit the size of these features to more
than 1 micron, the size related to the minimum scattering vector at
which we still observe the power law slope behavior. 
Voids of several hundred nanometers are  observed in the reconstructed
3D volume.

\begin{figure}[tbp]
	\includegraphics[width=0.45\textwidth]{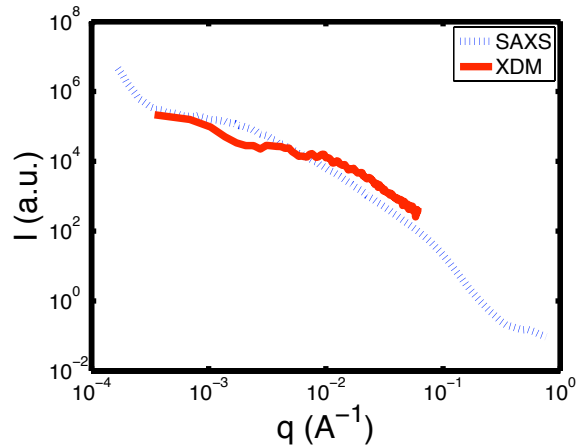}
\caption{
Small angle X-ray scattering data calculated from our 3D volume
reconstruction (XDM) compared to ultra small angle X-ray scattering
(SAXS) measurements on a similar batch of 100 mg/cc \ta2o5 aerogel. 
\label{fig:4} }
\end{figure}

We now turn our attention to understanding the mechanical properties of these
aerogels.  The strength and stiffness of three-dimensional foams are
often observed to scale as $\rho_a^m$, where $\rho_a$ is the density
of a structure divided by the density of its constituent members. In a
model proposed by Gibson and Ashby
\cite{gibson} the scaling exponent $m$, has limiting values of 1 where
deformation is axial, and 2 for structures that deform in
bending. Scaling relations in terms of $\rho_a$ assume some degree of
uniformity in the distribution of mass between the interconnecting
lattice ``beams'' and the nodes that define their intersections. These
assumptions appear satisfied for many foam-like structures at
densities down to 10\% or less ($\rho_a\lesssim 0.1$).  However the
strength of many low density aerogel samples of much lower density
(densities of less than 1\% or $\rho_a\ll 1$) is orders of magnitude
less than expected and scaling exponents between 3 and 4 are
frequently observed\cite{ma, pekala}.  Higher mass scaling exponents
are attributed to the presence\cite{ma2} of fragments disconnected
from the load-bearing backbone structure
\cite{pekala}, adding mass without carrying or transferring load. These
disconnected fragments are natural consequence of percolation: as a
structure approaches a critical density, more of its mass becomes
associated with branches disconnected from the backbone of the
lattice.

It is not clear, however, that such percolation models apply to these
aerogels
\cite{ma2, pirard}.  In one alternative model, heterogeneities such as
micron-sized holes produce spanning structures that could fail by buckling
\cite{pirard}. Yet another model proposes that diffusion-limited cluster
aggregation leads to fractal clusters (blobs) connected by thin beams (links)
\cite{ma2,meakin}.  The response of the ``blob and link'' architecture 
to compressive loading is simulated with
finite-element modeling, using the NIKE3D implicit finite-element
solver \cite{maker}. Material data for polycrystalline tantalum oxide
was used for the constitutive model with the modulus being 140 GPa.
For these simulations, opposing faces are loaded in compression, while
the unloaded cube faces are treated with mirror symmetry, restricting
boundary motion to the plane.  The average mass scaling exponent
obtained by this modelling was $m=3.6$, consistent with bulk data.
When the excess node mass was removed by computational thinning, the
scaling exponent reverted to $m=2$, indicating that the primary
deformation mechanism was bending of the interconnecting struts.

In this first experimental high-resolution view inside a foam, we see
a structure consisting of nodes connected by thin beams. These are
similar to the simulated fractal cluster aggregates and links derived
from a diffusion-limited cluster aggregation model.  This blob and
beam structure explains why these low density materials are weaker
than predicted, and explains the high mass scaling exponent observed
for this material.  Computational thinning of the measured structure
improves the strength-to-weight ratio by orders of magnitudes,
indicating that improvements in the strength could be obtained by
modifying the aerogel preparation conditions in an effort to
redistribute constituent material from nodes to interconnected struts.
Improvements in resolution using brighter light sources, shorter
wavelengths and larger detectors would enable resolving the cross
section of interconnecting struts, providing a full characterization
of not only these aerogels but also a range of engineered nanoscale
materials. The structural analysis we demonstrated here could be
applied to other porous materials and assist modelling percolation
problems such as oil and water in minerals\cite{percolation,bollobas}.
More generally the ability to image an unknown, isolated object in
three dimensions at high resolution, as demonstrated here, has  
the potential for a wide range of applications in material science, nanotechnology 
and biology at the cellular level.

\acknowledgments 
This work was supported by The U.S. Department
of Energy under Contract W-7405-Eng-48 to the University of
California, Lawrence Livermore National Laboratory; Projects 05-SI-003
and 05-ERD-003 from the Laboratory Directed Research and Development Program of LLNL;
the Advanced Light Source and the National Center for Electron
Microscopy, Lawrence Berkeley Lab, under DOE Contract
DE-AC02-05CH11231. 
JCHS and UW supported by DOE grant DE-FG03-02ER45996. The National Science
Foundation through the Center for Biophotonics, UC Davis, under
Cooperative Agreement No. PHY0120999.  Use of the Advanced Photon Source was supported by the U.S. Department of Energy, Office of Science, Office of Basic Energy Sciences, under Contract No.W-31-109-ENG-38.
The authors have no competing financial interests.


\section*{
Supplemental online material
}

\paragraph*{Experimental equipment and methods:}
The Aerogel sample was imaged at beamline 9.0.1 \cite{howells} of the
Advanced Light Source (ALS), Berkeley, using the diffraction apparatus
of Stony Brook University described in detail elsewhere
\cite{beetz}. The sample is coherently illuminated by 750eV x-rays
(1.65 nm wavelength) from an undulator source using a zone plate
monochromator. A 5 $\mu$m pinhole at the zone plate focus selects a
transversely coherent patch of the illuminating beam and also selects
the wavelength with a spectral resolution of $\lambda/\Delta \lambda =
1000$.  The finite spectral resolution limits image width to 1000
resolution elements in each direction \cite{Chapman_Pyramid}.  The
5-$\mu$m-diameter monochromator exit pinhole also selects a
transversely spatial coherent patch of the beam.

The sample is located 20 mm downstream of this pinhole and is rotated
from 69\degr to 64\degr in 1\degr increments perpendicular to the
X-ray beam to obtain diffraction patterns for different projections
through the object, the missing angle range corresponding to sample
tilts where the beam is obscured by the slotted window. A moveable
beamstop blocks the undiffracted direct beam from impinging on the
CCD, as shown in Figure 1 of the main text, and is translated in
separate exposures to minimise the region of missing data at low
spatial frequencies.  Diffraction from the sample is measured using a
water-cooled Princeton Instruments PI-MTE 1300 direct detection
in-vacuum CCD camera located 164mm behind the sample. The CCD chip
itself is composed of a 1300x1320 array of 20 $\mu$m pixels and is
cooled to -45\degr C to reduce dark noise.  We selected sub-arrays of
$1200 \times 1200$ elements, centered on the location of the zero
spatial frequency (direct beam).  At these CCD and wavelength settings
we have a real-space sampling interval in $x$ and $y$ of $\Delta x =
11.3 nm$ (in the small-angle approximation) and a field width of $w =
13.6 \mu m$.  Because the intensity across the diffraction pattern
varies by orders of magnitude, multiple exposures were taken at each
rotation and summed together to increase the CCD dynamic range
\cite{Chapman_Pyramid}.

\paragraph*{Phase retrieval:}
For each view of the sample, the measured diffraction intensities are
proportional to the modulus squared of the Fourier transform of the
wave exiting the object.  However the diffraction pattern on its own
contains incomplete information about the object - obtaining an image
of the object requires knowledge of the optical phase of the
diffraction pattern.

If the diffraction pattern intensities are sampled finely enough it is
possible to solve for the diffraction pattern phases and, thus,
produce an image of the object in real space
\cite{fienup,marchesini}. The feasibility of this technique for
reconstructing an image of the sample from its diffraction pattern has
been well demonstrated in many X-ray diffraction experiments
\cite{miao,miao:prl,shapiro,Chapman_Pyramid,pfeifer}.

The individual measured diffraction patterns consist of 2D CCD images,
which in turn correspond to Ewald sphere sections through the 3D
diffraction volume.  The Ewald sphere geometry is completely defined
by the known experimental geometry (pixel size, CCD distance,
wavelength, sample rotation, etc.).  It is therefore possible to
uniquely map the locations of each pixel of each diffraction patterns
into a corresponding location in the 3D diffraction volume.  The
transform to perform this mapping is described in detail elsewhere
\cite{Chapman_Pyramid}.  Here, we used a simple gridding procedure
that placed individual 2D intensities in the nearest corresponding 3D
voxel, although we note that more sophisticated interpolation
algorithms could be applied.  All other voxels in the data set were
left blank, denoting regions of unknown data.  This Ewald sphere
mapping is performed such that the highest resolution data collected
by the selected $1200 \times 1200$ sub-arrays is fully captured in the
3D diffraction volume.  Pixels where the Fourier space data was not
known were allowed to be unconstrained in both Fourier space phase and
magnitude during the reconstruction process.  We note that Ewald
sphere shells have a maximum separation of no more than 4 pixels at
the edge of the diffraction volume, smaller than the observed speckle
size in our data.  Because the measured diffraction pattern is
invariant with respect to translation of the object and reconstruction
is performed directly on the three dimensional diffraction volume,
there is no need to align individual 2D projections with respect to
one another before reconstructing the 3D image.  Instead, it is
necessary to determine the centre of each diffraction pattern before
assembling the 3D diffraction volume.

To obtain the real-space object structure from the diffraction volume,
iterative phase retrieval was performed using a 3D implementation of
the HIO algorithm
\cite{fienup} with feedback parameter $\beta=0.9$ for 1200 iterations,
 followed by the
RAAR algorithm \cite{luke} through to iteration 3000 also with
feedback parameter of $\beta=0.9$.  Values in the object
reconstruction were allowed to be complex valued, and phase retrieval
was performed based on the measured diffraction data alone.  Due to
both the size of the reconstruction mesh and the large number of 3D
Fourier transforms required to perform iterative phase retrieval a
cluster computing solution was used to process the
data. Reconstructions were performed on a 16-node 2.0GHz dual
processor Macintosh G5 X-serve cluster (32-CPU in total) with
Infiniband interconnects and 4GB memory per node.  For optimum Fourier
transform speed we used the dist\_fft distributed fast Fourier
transform library from Apple Computer.
volume 
using either
2 Fourier transforms per iteration
This diffraction volume supports a field of view of 5.8 $\mu m$ with a
real-space 
3D reconstruction for this diffraction volume required 3000 iterations
of the phase retrieval loop and took 2.5 hrs to complete.  Methods for
handling the missing angles and limited number of views were identical
to those previously described elsewhere, and the spatial resolution of
our result is necessarily band-limited by the CCD numerical aperture
\cite{Chapman_Pyramid}.

\paragraph*{Image resolution:}
Critical estimation of the reconstruction resolution is important and
can be performed using a variety of conventions adopted by different
research groups.  The spatial resolution of our imaging system is
ultimately limited by the subtended numerical aperture, describing the
highest spatial frequency in the object which could theoretically be
imaged: finer resolution detail is simply not captured by the
detector, therefore can not be physically measured.  For our imaging
geometry (750eV, 20 $\mu m$ pixels, $1200 \times 1200$ pixel
sub-array, and 164 mm CCD distance) we have an upper limit on spatial
resolution determined by the subtended numerical aperture - namely a
resolution of 11.3 nm at the edge of the CCD, decreasing to 8 nm in
directions towards the corners of the CCD chip. Resolution is degraded
in the longitudinal direction due to the wedge of missing data, as
previously observed \cite{Chapman_Pyramid}.

Achieving this resolution assumes a perfect detector, adequate
measured signal out to the resolution limit, and a perfect phase
retrieval algorithm.  In diffraction microscopy intensities are
sampled not in a focal plane of a lens but in the plane of the lens
itself, therefore it is necessary to have sufficient photons measured
at the claimed spatial resolution limit in order to justify any
resolution claims.  Crystallographers have, by convention, adopted a
signal-to-noise ratio (SNR) of 3 as the threshold of resolution,
measured as the SNR in a Bragg peak compared to the background noise
level.  The equivalent measure in diffraction microscopy of a
continuous object, where there are no discrete Bragg peaks, is the SNR
of the speckles relative to the background noise level.  For our data,
the "SNR $>$ 3" criterion is satisfied out to the edge of the CCD
camera in each individual diffraction image, where we have a mean of
18 photons/pixel signal above background levels (SNR $>$ 4).

It is well known that sufficient angles must be measured so that the
spacing between Ewald spheres in reciprocal space is less than the
typical spacing between Bragg peaks in order to avoid artifacts from
interpolation.  In the case of continuous diffraction patterns, the
angular sampling must be at least as fine as individual speckle size
in the diffraction data.  For our data set, the object width is
$\approx 3 \mu m$, thus the oversampling ratio for this case is
greater than 4 $(s = 4.5)$.  With 1 degree sampling, the spacing
between measured planes is at most 4 pixels at the very edge of the
diffraction volume, implying that each speckle is at sampled at least
once in our data.  Crowther's criterion, on the other hand, dictates
that the critical sampling of diffraction intensities ($s=2$) when
\begin{equation}
  \Delta \phi = \Delta q/q_\mathrm{max} = \Delta x/D.
\end{equation}
leading to a requirement for 0.25$^\circ$ separation between planes in
order to adequately sample the finest features.  In practice we
collect diffraction data with angular increments that are 2--4 times
larger than the Crowther criterion. In the process of phase retrieval
we additionally recover both the amplitudes and phases of the missing
data between the Ewald surfaces, including those in a large gap
resulting from a limited range (usually $\pm 70^\circ$) of rotation
angles, data blocked by a beam-stop, and the missing ``cone'' of data
resulting from rotating the sample about a single axis.  The
quantitative effects of this procedure on image resolution was studied
in detail in \cite{Chapman_Pyramid}, to which we refer readers for a
detailed discussion of the subject.  The effect of this missing wedge
of missing data corresponding to inaccessible sample rotation angles
yields lower resolution in the $z$ direction as previously observed
\cite{Chapman_Pyramid}.

In diffraction imaging, reconstruction algorithms and techniques
perform the role typically ascribed to lenses in an imaging system. It
is therefore important to assess performance of the phase retrieval
process in addition to the photon detection system for stability and
uniquesness of the final solution.  The performance of our imaging
technique could, in principle, be quantified in Fourier space by
measuring the modulation transfer function (MTF) of the imaging system
as a whole.  For the numerical reconstruction technique used here this
MTF would encapsulate resolution limits due to signal-to-noise, data
alignment and regions of missing data, as well as algorithm stability
and uniqueness.  The phase retrieval process recovers the diffraction
phases with a limited accuracy, due to factors including SNR of the
diffraction amplitudes, missing data, the inconsistency of
constraints, and systematic errors in the data (such as errors in
interpolation).  The MTF itself can be calculated only if the original
object is known, and is therefore not applicable to evaluating the
resolution of an unknown object.  However, we can compute the
stability of the phase retrieval process itself by comparing the mean
reconstructed Fourier space intensities with our measured data.  An
equivalent MTF for phase retrieval can be determined by calculating
the phase retrieval transfer function (PRTF), which represents the
effective transfer function of our numerical imaging system.  Details
of the PRTF calculation have been previously in detail elsewhere
\cite{shapiro, Chapman_Pyramid}.  Where the phases are consistently
retrieved to the same value, the squared modulus of the average will
be equal to the constrained modulus, and the ratio will be unity.
Where the phases are random and completely uncorrelated, the average
will approach zero.  Thus, the ratio is effectively a transfer
function for the phase retrieval process, and the average image is the
best estimate of the image: spatial frequencies are weighted by the
confidence in which their phases are known \cite{shapiro}.  A
conservative estimate of the resolution is given by the frequency at
which the PRTF reaches a value of 0.5. We calculated the PRTF for our
reconstruction, and it remains above a value of 0.5 right to the edge
of our measured data, indicating that we indeed obtained a diffraction
limited image of the sample.

Finally, we note that imaging system resolution is typically evaluated
using a known object rather than an unknown object.  The reasons are
obvious: with a known object it is possible to quantitatively compare
the initial object with the image and thereby determine the resolution
of the imaging system, whereas for an unknown object there is
necessarily doubt as to whether it is the object or the imaging system
which is imposing limits on spatial resolution.  System resolution is
typically evaluated by first imaging a known object to determine
system resolution, and then imaging subsequent objects under identical
conditions: if the imaging process is assumed to be deterministic, the
achievable resolution for the two imaging systems should be the same.
We followed this standard approach by first imaging a known test
object and analysing the spatial resolution in detail, then applying
the same imaging methods to unknown samples.  Our discussion of system
resolution were presented in detail elsewhere \cite{Chapman_Pyramid},
demonstrating that the imaging technique applied here faithfully
reproduces a known test object up to the stated resolution limits.
Here, we apply the same techniques to a new, unknown object, and once
again arrive at a similar estimate of system resolution.  Unlike the
gold ball pyramid, the aerogel sample imaged here is of unknown
structure and therefore limited use in performing a detailed analysis
of system resolution.


\begin{figure}
	\includegraphics[width=0.48\textwidth]{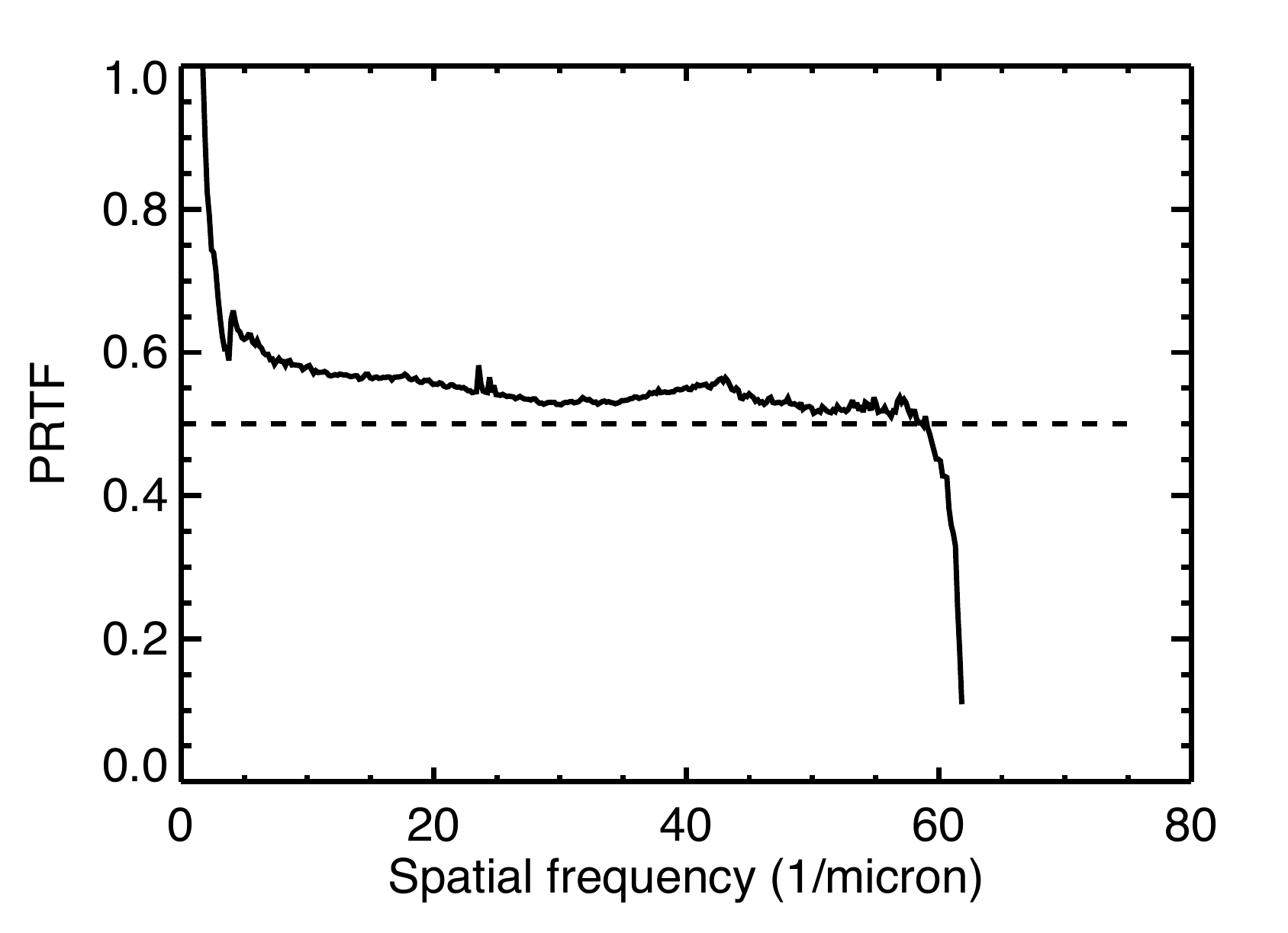}
\caption{
Phase retrieval transfer function (PRTF) averaged over shells of
constant $u$ for the entire 3D data volume.  The PRTF remains above
0.5 to the edge of measured data, indicating that our phase retrieval
produces a diffraction limited reconstruction to a resolution of 11.3
nm \cite{Chapman_Pyramid}, corresponding to the spatial frequency at
the edge of the CCD.  Indeed the PRTF remains above 0.5 out to 8 nm,
which is near the diagonal corner of the CCD.  }
\label{PRTF}
\end{figure}

\paragraph*{Finite element analysis:}
For finite element analysis, a subset was taken from the interior of
the reconstructed particle near the center to produce a finite box of
material. The mesh was generated by assuming that any voxel partially
occupied at least half-way (as determined by the voxel value) was
solid material, followed by a threshold to generate the mesh. The
voxel size was 8.9 nm and the cube was 94 voxels on a side. The
assigned modulus and hardness were taken from the literature as 140
GPa and 5.3 GPa, respectively. The volume fraction was estimated as
the number of elements (74150) divided by the volume of the cube
(943), which is ~0.089. Using mirror boundary conditions, the cube was
compressed parallel to each of the three faces to a strain of 1
stiffness is strongly dependent upon the orientation. The estimated
stiffness in the X, Y, and Z directions were 25.0, 11.6, and 16.7 MPa,
respectively. Based on the calculated volume fraction the exponents of
the modulus vs fraction power laws for each direction are 2.62, 3.89,
and 3.74, for XYZ respectively. The highest stresses were recorded in
the X compressed sample was ~2.3 GPa at 1\% compression.  The
resulting stress-strain response along with the associated stiffness
is plotted in Figure 4.

\paragraph*{USAXS data:}
USAXS data was acquired at beamline 33-ID at the Advanced Photon
Source, Argonne National Laboratory \cite{USAXS_Ilavsky1,
USAXS_Ilavsky2}. The end station consists of a Bonse-Hart camera,
which can measure scattering vectors (q) from about 0.0015 to 10
nm-1. The energy of the incident x-rays in the scattering experiments
was 9.7 keV below the tantalum L3 edge. Scattering data was processed
using the codes developed for use on this USAXS instrument, and
included absolute scattering intensity calibration and slit desmearing
\cite{USAXS_Ilavsky3}.
In the Ta$_2$O$_5$ aerogel two distinct power-law regimes are observed
with a exponent of Ð4 in the Porod region, and approxiately -2 in the
form factor region. The cross over points between the changing slopes
occur at $q_1 \approx 0.1 \AA^{-1}$ and $q_2 \approx 0.09 \AA^{-1}$
with an associated radii of gyration of $Rg_1 \approx 20 \AA$ and
$Rg_2 \approx 140 \AA$ respectively.  Values of the power law
exponents and radii of gyration are obtained by standard Porod and
Guinier analysis. For aerogels a fractal analysis is commonly used to
interpret the scattering data. In the fractal model $Rg_1$ and $Rg_2$
are related to the mean particle diameter and the correlation range
(pore diameter or fractal size). A weakness of this analysis is that
the relation between $Rg$ and size or correlation range depends
strongly on the shapes and size distribution of the scatters. If we
assume roughly spherical particles $Rg_1$ translates into a mean
diameter of ~ 5 nm. This agrees favorable with TEM analysis of the
tantala aerogel which images particles on the order of 3-5 nm in
diameter particularly given that the Guinier analysis overestimates
the mean diameter in a polydispersed systems\cite{Hasmy}.  We note
that the scattering data for the tantala aerogel is similar to that
measured from a silica aerogel
\cite{Schaefer}.

\paragraph*{USAXS comparison:}
Calculated SAXS data for the Aerogel foam we imaged was generated by
extracting a cube of 128 voxels on a side consisting of only the
Aerogel portion of the reconstructed volume. This was Fourier
transformed to obtain scattering from the Aerogel structure alone, and
radially averaged to obtain SAXS data corresponding to three
orthoigonal views through the sample.  Scattering angle q-vectors were
calculated from the known reconstruction voxel size, and an arbitrary
scale factor was applied to the intensity for comparison to the APS
data (this shifts the plot vertically on Figure 4 but does not change
the gradient on a log plot).  We note that SAXS analysis performed on
both the entire reconstructed volume, and on the raw diffraction data
collected at beamline 9.0.1, did not agree with the APS USAXS data.
We attribute this to scattering from the membrane and surrounding
particles, as visible in the 3D reconstruction, and which could be
avoided by extracting a sub-cube from the 3D data as described above.

\end{document}